\documentclass[cite]{PoS}
\usepackage{xspace}
\newcommand{\aof}{A\,0535+26\xspace}
\newcommand{\hde}{HDE\,245770\xspace}

\newcommand{\inte}{\textsl{INTEGRAL}\xspace}
\newcommand{\xte}{\textsl{RXTE}\xspace}

\newcommand{\pca}{\textsl{PCA}\xspace}

\newcommand{\hexte}{\textsl{HEXTE}\xspace}
\newcommand{\ibisgr}{\textsl{IBIS (ISGRI)}\xspace}
\newcommand{\ibis}{\textsl{IBIS}\xspace}

\newcommand{\swift}{\textsl{Swift}\xspace}
\newcommand{\bat}{\textsl{BAT}\xspace}

\newcommand{\suzaku}{\textsl{Suzaku}\xspace}

\newcommand{\hexe}{\textsl{HEXE}\xspace}
\newcommand{\osse}{\textsl{OSSE}\xspace}

\title{Peculiar outburst of \aof observed with \inte, \xte and \suzaku}

\ShortTitle{Peculiar outburst of \aof observed with \inte, \xte and \suzaku}

\author{\speaker{I.~Caballero}$^a$, 
  K.~Pottschmidt $^b$,
  L.~Barrag\'{a}n $^c$,
  C.~Ferrigno $^d$,
  P.~Kretschmar $^e$,
  S.~Suchy $^f$,
  J.~Wilms $^c$,
  A.~Santangelo $^g$,
  I.~Kreykenbohm $^c$,
  R.~Rothschild $^f$,
  D.~Klochkov $^g$,
  R.~Staubert $^g$,
  M.~H.~Finger $^h$,
  A.~Camero-Arranz $^{h,i,j}$,
  K.~Makishima $^{k,l}$,
  T.~Mihara $^{l}$,
  M.~Nakajima $^m$,
  T.~Enoto $^k$,
  W.~Iwakiri $^n$,
  Y.~Terada $^n$\\
  \llap{$^a$}  CEA Saclay, DSM/IRFU/SAp --UMR AIM (7158) 
  CNRS/CEA/Universit\'{e} P.Diderot --F-91191 Gif sur Yvette France\\
  \llap{$^b$} CRESST, UMBC, MD 21250 /NASA
  GSFC, Code 661, Greenbelt, MD 20771, USA\\
  \llap{$^c$}  Dr. Karl Remeis-Sternwarte -- FAU Erlangen-N\"{u}rnberg, 96049 Bamberg, Germany\\
  \llap{$^d$} ISDC Data Centre for Astrophysics, 1290 Versoix, Switzerland \\
  \llap{$^e$} ISOC, European Space Astronomy Centre, ESA, 28691 Villanueva de la Ca\~{n}ada, Madrid, Spain\\
  \llap{$^f$}Center for Astrophysics and Space Science, UCSD,  La Jolla, CA, USA\\
\llap{$^g$}Institut f\"{u}r Astronomie und Astrophysik, Sand 1, D-72076 T\"{u}bingen, Germany \\
  \llap{$^h$}NSSTC, 320 Sparkman Drive NW, Huntsville, AL 35805 USA\\
  \llap{$^i$}Fundaci\'{o}n Espa\~{n}ola de Ciencia y Tecnolog\'{i}a / MICINN,  Madrid, Spain\\
  \llap{$^j$}MICINN (Ministerio de Ciencia e Innovaci\'{o}n), C/Albacete, 5, 28027, Madrid, Espa\~{n}a\\
  \llap{$^k$}University of Tokyo, Japan\\
  \llap{$^l$}RIKEN, Japan\\
  \llap{$^m$}Nihon University, Japan\\
  \llap{$^n$}Saitama University, Japan\\
  E-mail: \email{isabel.caballero@cea.fr}}

\abstract{A normal outburst of the Be/X-ray binary system A0535+26 has taken 
place in August 2009. It is the fourth in a series of normal outbursts
that have occured around the periastron passage of the source, but  
is unusual by starting at an earlier orbital phase and by presenting 
a peculiar double-peaked light curve. 
A first "flare" (lasting about 9 days from MJD 55043 on)  
reached a flux of 440 mCrab. The flux then decreased to less than 220 mCrab, 
and increased again reaching 440 mCrab around the periastron at MJD 55057. 
Target of Opportunity observations have been performed with INTEGRAL, 
RXTE and Suzaku. First results of these observations are presented, 
with special emphasis on the cyclotron lines present in the X-ray spectrum 
of the source, as well as in the pulse period and energy dependent 
pulse profiles of the source.}

\FullConference{The Extreme sky: Sampling the Universe above 10 keV - extremesky2009,\\
		October 13-17, 2009\\
		Otranto (Lecce) Italy}

\begin{document}
\section{Introduction}
\aof is a Be/X-ray binary system, discovered by Ariel V
during a giant outburst in 1975 \cite{rosenberg75}. The binary system 
consists of the pulsating neutron star \aof and the optical companion \hde \cite{bartolini78}. 
It lies in an eccentric orbit of $e=0.47$ with an orbital 
period of $P_{\mathrm{orb}}\sim111\,$ days \cite{finger06}. 
The distance to the system is $d=1.8\pm0.6\,$kpc \cite{giangrande80}, later confirmed by 
e.g. \cite{steele98}.  An extensive review 
is given in \cite{giovanelli92}.  Since its discovery, five \footnote{At the moment of writing, 
December 2009, a new giant outburst of the source is taking place, \cite{wilson09}}
giant outbursts have been detected, in October 1980 \cite{nagase82}, June 
1983 \cite{sembay90}, March/April 1989 \cite{makino89}, February 1994 
\cite{finger94_1} and May/June 2005 \cite{tueller05}. The source presents 
cyclotron lines in its X-ray spectrum. Cyclotron lines are very powerful 
tools, since they are the only direct way to determine the magnetic field of 
a neutron star. \aof shows two cyclotron lines at $\sim$45\,keV (fundamental) 
and $\sim$100\, keV (first harmonic). These lines were observed in 1994 
with \hexe and \osse (\cite{kend94}, \cite{grove95}), and have later 
been confirmed with \inte, \xte \cite{kretschmar05} and \suzaku 
\cite{terada06}. From the cyclotron lines, the magnetic field is inferred 
to be B$\sim$4$\times10^{12}$\,G \cite{caballero07}.

\section{Renewed activity since 2005}
\begin{figure}
\begin{center}
\includegraphics[angle=90,width=0.7\textwidth]{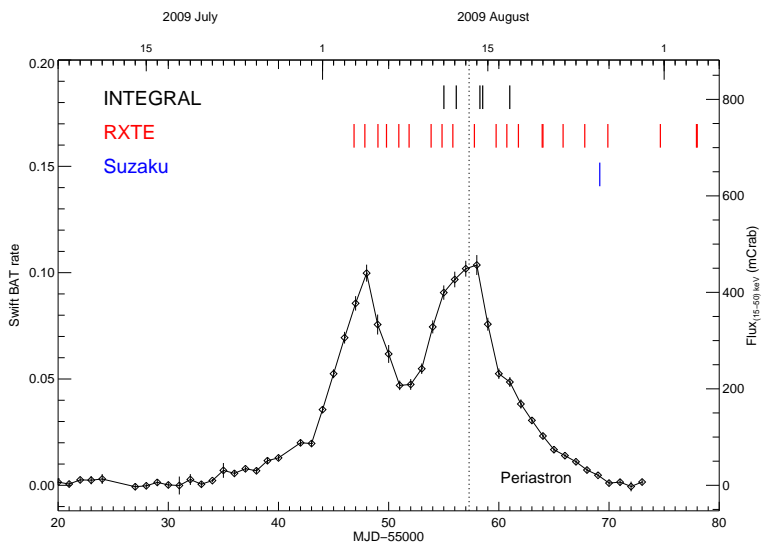}
\end{center}
\caption{\label{fig:lc}
\bat 15-50\,keV light curve of the peculiar August 2009 outburst of \aof. The 
times of our \inte, \xte and \suzaku TOO observations are indicated, as well 
as the periastron passage.}
\end{figure} 

As mentioned above, after more than 11 years of quiescence the source showed 
a giant outburst in 2005, which unfortunately could not be observed due to 
Sun constraints. Two subsequent normal outbursts took place  in 
August/September and December 2005.  Flaring activity was discovered during the 
August/September outburst, as well as associated changes in the cyclotron 
line energy and energy dependent pulse profiles. The cyclotron line was 
measured at $\sim52\,$keV during one of the flares, significantly higher 
than during the rest of the outburst \cite{caballero08_1}. These
changes were interpreted in terms of magnetospheric instabilities developing 
at the onset of the accretion \cite{postnov08}.
Since then, the source has shown several normal outbursts associated with 
the periastron. 

The one from August 2009 \footnote{The outburst subject of this work 
preceeds the December 2009 giant outburst, which is also taking place around periastron, 
about one orbital period after the August 2009 outburst \cite{caballero09_c}. } is the fourth 
in a series of 
normal outbursts, but is peculiar,  starting at an earlier orbital phase and 
showing an unusual double-peaked light curve. The \swift-\bat light curve 
(see Fig.~\ref{fig:lc})  shows a first ``flare'' that lasted about 9 days
from MJD 55043, about 14 days before the periastron, reaching a flux of 
~0.1 counts$/$s$/$cm$^2$ (~440 mCrab) in the 15-50 keV Swift/BAT lightcurve 
before decreasing again to less than 0.05 counts$/$s$/$cm$^2$. The flux then 
rose again reaching ~0.1 counts$/$s$/$cm$^2$ (~440 mCrab) in the Swift/BAT at 
MJD 55057 around the periastron (\cite{finger09}, \cite{caballero09_b}).
This new outburst has led to our \inte, \xte and \suzaku TOO observations 
being triggered. The times of the observations are shown in Fig.~\ref{fig:lc}.  
A similar double-peaked normal outburst was already observed for \aof prior to the
1994 giant outburst \cite{finger96}. 
The August 2009 outburst might be comparable to the 'noisy' or 'anomalous' outbursts discussed in 
\cite{giovanelli92}, although the different sampling makes this difficult to assess.


\section{First results}

\begin{figure}
\centerline{
\includegraphics[angle=90,width=0.6\textwidth]{%
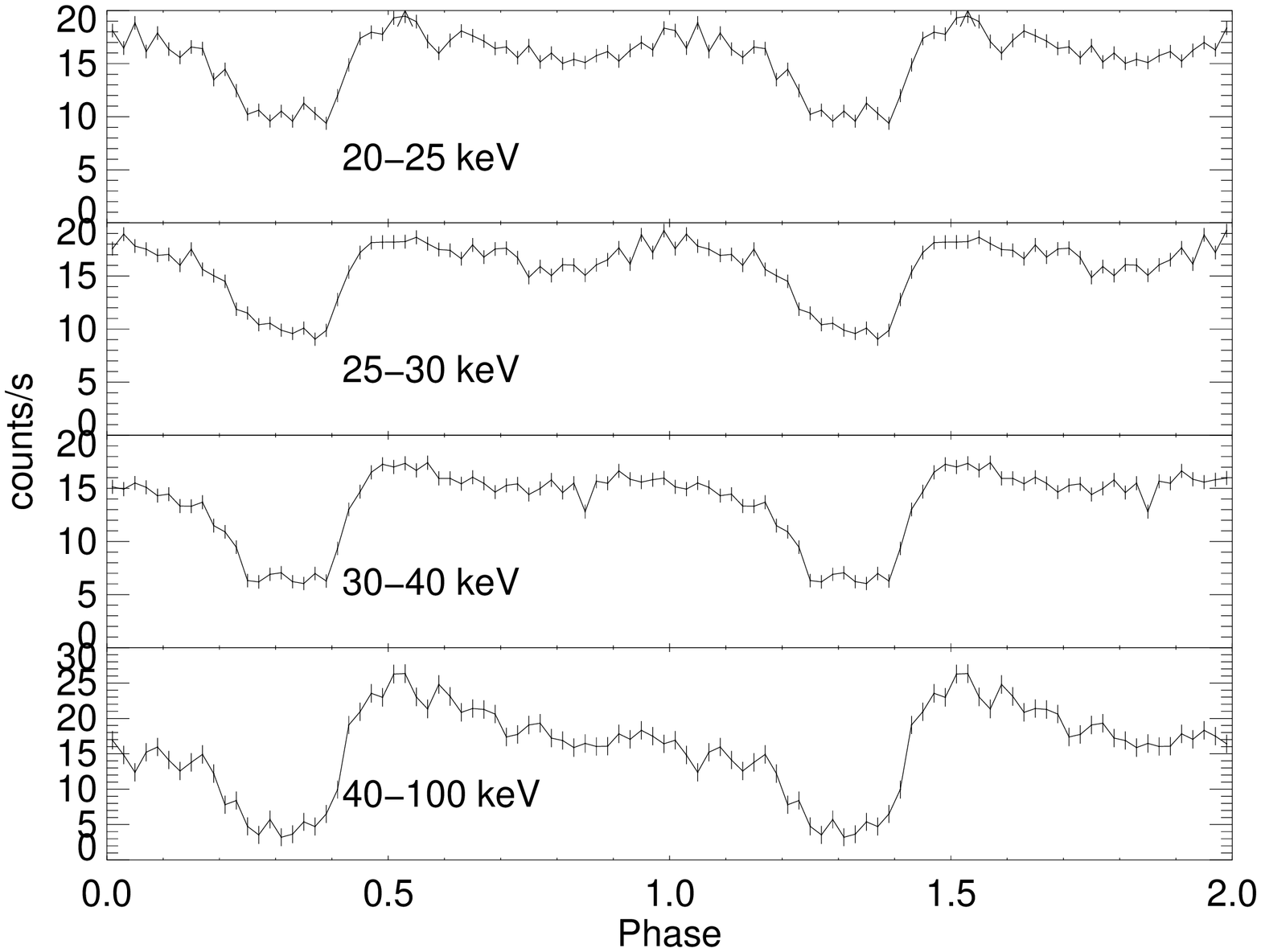}
}
\caption{\label{fig:pp} Energy dependent pulse profiles of \aof obtained 
with \inte \ibisgr.}
\end{figure}

First results reveal pulsations of the neutron star with a spin frequency 
of $\nu=9.6608[2]10^{-3}$\,Hz and a derivative of 
$\dot{\nu}=1.2[6]\times10^{-12}$\,Hz$/$s or a spin period of 
$P=103.511[2]\,$s and period derivative 
$\dot{P}=-1.2[6]\times10^{-8}\,$s\,/s, measured at MJD 55054.995 
using \inte \ibis data. The orbital correction was performed 
using the ephemeris from \cite{finger06}. 
Energy dependent pulse profiles of one of our \inte 
observations are shown in Fig.~\ref{fig:pp}. The pulse profiles show a 
double-peaked structure at lower energies, with one of the peaks being 
reduced above $\sim$45\,keV. These pulse profiles are very similar to 
pulse profiles of the source observed in the past (see e.g. \cite{kend94}, 
\cite{finger96}, \cite{caballero07}). 

\begin{figure}
\begin{center}
\centerline{%
\includegraphics[angle=90,height=0.4\textwidth]{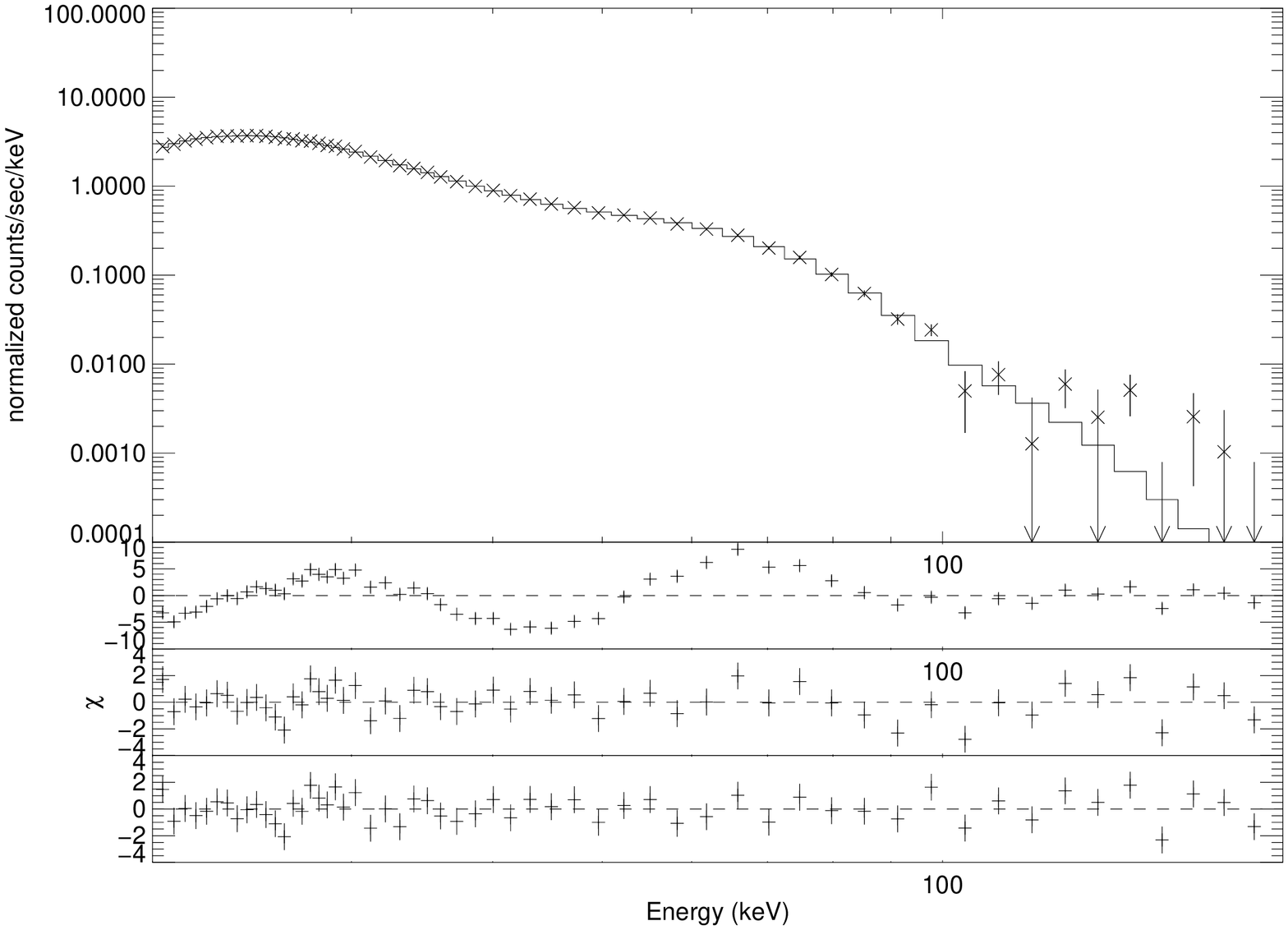}
\includegraphics[angle=0,width=0.45\textwidth]{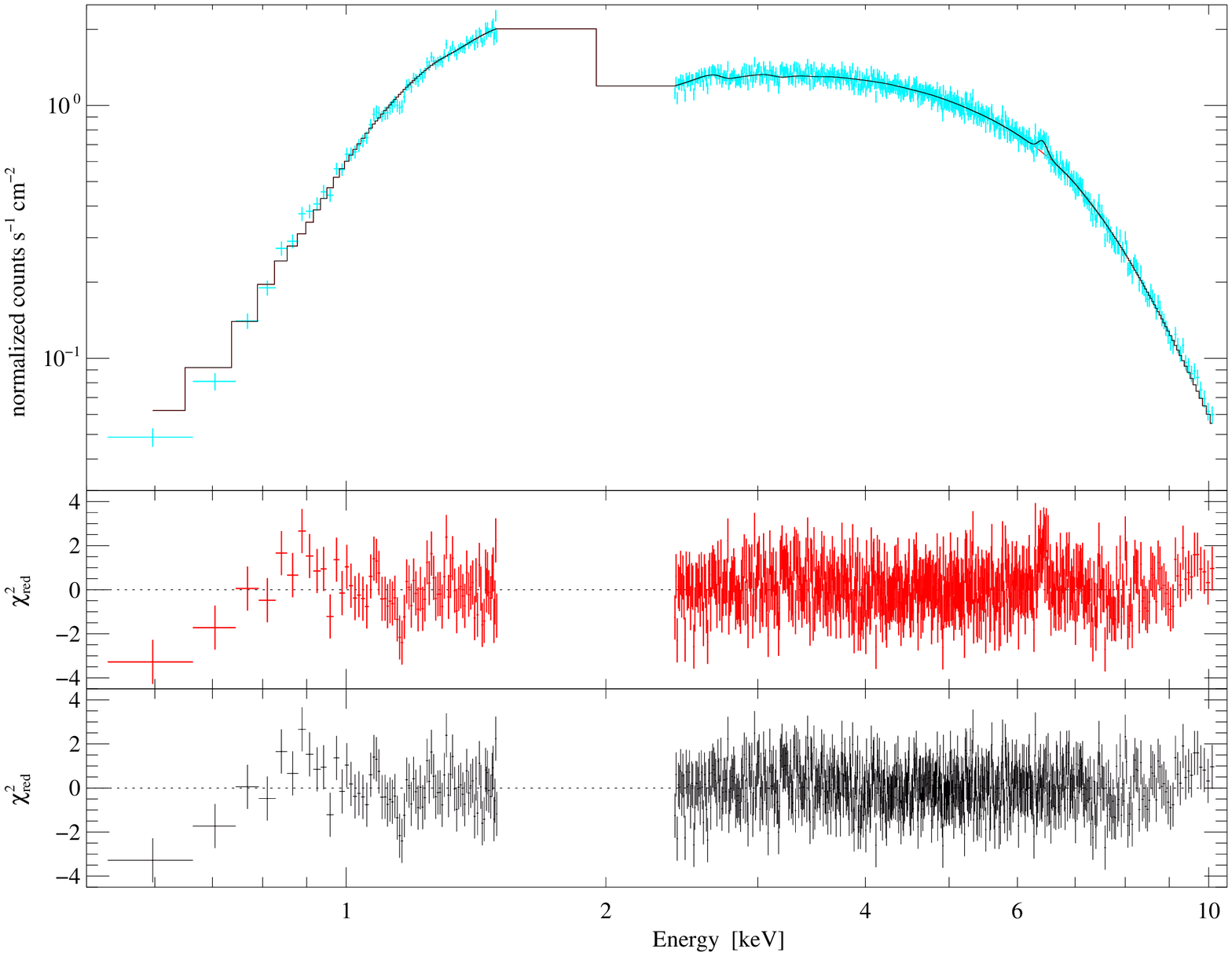}}
\caption{\label{fig:spec1} Left: \inte \ibisgr spectrum of \aof. The upper panel 
shows the data and model, and the lower panels show the residuals of a fit 
including no absorption lines, one and two lines at $\sim45$ and 
$\sim100\,$keV in the model. Right: \suzaku XIS spectrum of \aof. 
The upper panel shows the data and model, the middle and lower panel show the 
residuals of a fit without and with an Fe fluorescence line in the model.}
\end{center}
\end{figure} 

The X-ray continuum spectrum is modeled with an absorbed cutoff powerlaw. 
Absorption-like features are seen in both \xte and \inte data, interpreted 
as cyclotron lines. We modeled the cyclotron lines using Gaussian lines in 
absorption \cite{coburn02}.  As an example, an \ibis spectrum is shown in 
Fig.~\ref{fig:spec1}. From 
our first results, the best fit value for the fundamental cyclotron line 
is $E_{\mathrm{cyc}}=45.4^{+0.7}_{-0.6}$\,keV  for \ibisgr and 
$E_{\mathrm{cyc}}=44.5^{+0.8}_{-1.1}$ \,keV for \hexte. The \ibis data require a 
second line fixed at 102\,keV.  These values are in agreement with 
the values obtained in past outbursts (\cite{kend94}, 
\cite{terada06}, \cite{caballero07}).
First results from the \suzaku analysis are given in Fig.~\ref{fig:spec1} (right), 
in which a XIS spectrum is shown. The 1.5-2.4\,keV energy range is currently 
ignored due to calibration issues (ongoing work).   
We measure an Fe fluorescence line at $E=6.42^{+0.03}_{-0.02}\,$keV and 
an absorption column density of 
$N_{\mathrm{H}}=0.48\pm0.02\times 10^{22}\mathrm{cm}^2$. 

\begin{figure}
\begin{center}
\includegraphics[width=0.7\textwidth]{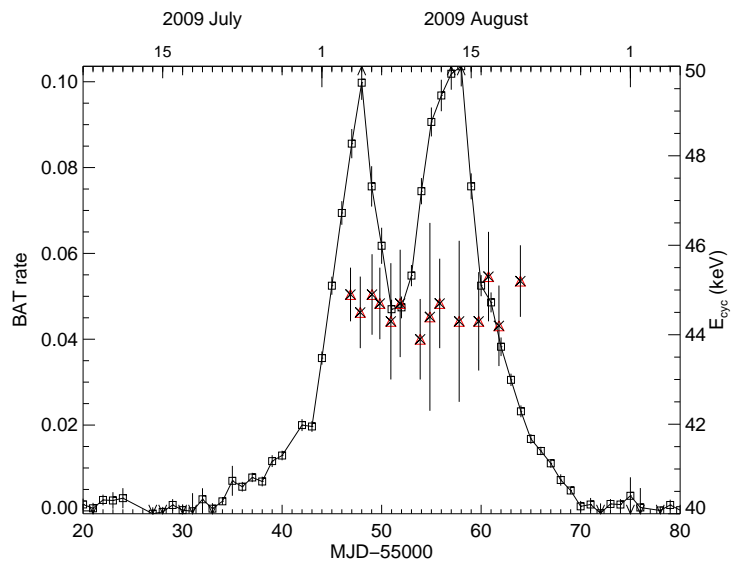}
\end{center}
\caption{\label{fig:xte}%
\swift-\bat light curve during the 2009 outburst (left axis, black squares) and 
cyclotron line energy as measured with \xte (right axis, red triangles).}
\end{figure}

Making use of the \xte observations (\pca and \hexte data)  
we have studied the evolution of the 
cyclotron line energy during the outburst. Preliminary results are shown in 
Fig.~\ref{fig:xte}. There seems to be no significant variation of the 
fundamental cyclotron line energy during the outburst. The analysis of the 
last observations at lower luminosity is still ongoing.

\vspace*{-5mm}

\section{Conclusions and outlook}
Work is ongoing to analyze the \inte, \xte and \suzaku observations. 
One main goal is to study the dependence of the cyclotron line energy with 
the luminosity in order to probe changes of the accretion column height 
for different accretion rates of the source (see \cite{basko_sunyaev_76} 
and \cite{mihara95}, \cite{tsygankov06}, \cite{staubert07} and 
references therein). First results show no evidence for a change in the 
cyclotron line energy with the luminosity, suggesting that the line forming 
region is not changing with the luminosity of the system. This result is 
in agreement with what was found in the 2005 outburst \cite{caballero07}. 
We will also study the energy dependent pulse profiles during the outburst. 
Work is also ongoing to study the pulse period and spin-up of the source, evidence 
for an accretion disk. The presence of a temporary accretion disk around the periastron could
be detected also in optical as discussed by \cite{giovanelli07}. 
We will also search for possible QPOs, as already observed in the source in 1994 \cite{finger96}.

We are currently monitoring the current December 2009 giant outburst with 
\xte. We plan to perform a detailed spectral and timing analysis of the new 
outburst, that will be related and compared with the findings of the 
August 2009 outburst. 

\section*{Acknowledgments}
IC acknowledges support from the French Space Agency CNES. 

\end{document}